\documentclass[12pt]{iopart}
\usepackage{graphicx}
\begin{document}

\title[Power counting and renormalization group invariance in the SKM]{Power counting and renormalization group invariance in the subtracted kernel method for the two-nucleon system}

\author{S. Szpigel $^1$ and V. S. Tim\'oteo $^2$}
\address{$^1$ Faculdade de Computa\c{c}\~ao e Inform\'atica, Universidade Presbiteriana Mackenzie,
01302-907, S\~ao Paulo, SP, Brazil}
\address{$^2$ Faculdade de Tecnologia, Universidade Estadual de Campinas - UNICAMP,
13484-332, Limeira, SP, Brazil}
\ead{szpigel@mackenzie.br and varese@ft.unicamp.br}

\begin{abstract}
We apply the subtracted kernel method (SKM), a renormalization approach based on recursive multiple subtractions performed in the kernel of the scattering equation, to the chiral nucleon-nucleon ($NN$) interactions up to next-to-next-to-leading-order ($NNLO$). We evaluate the phase-shifts in the $^1 S_0$ channel at each order in Weinberg's power counting scheme and in a modified power counting scheme which yields a systematic power-law improvement. We also explicitly demonstrate that the SKM procedure is renormalization group invariant under the change of the subtraction scale through a non-relativistic Callan-Symanzik flow equation for the evolution of the renormalized $NN$ interactions.
\end{abstract}


\section{Introduction.}

The issue of the non-perturbative renormalization of nucleon-nucleon ($NN$) interactions in chiral effective field theory (ChEFT) has been intensively investigated by many authors (for detailed reviews, see e.g. Refs.~\cite{bedaque1,epelbaum1a,machleidt1,epelbaum1b}), generating a great deal of discussion and debate regarding the consistency of the approach originally proposed by Weinberg \cite{weinberg1,weinberg2,weinberg3}. The standard procedure for the non-perturbative renormalization of $NN$ interactions in the context of Weinberg's approach to ChEFT can be divided in two steps \cite{epelbaum1a}. In the first step, one has to solve a regularized Lippmann-Schwinger (LS) equation for the scattering amplitude by iterating the effective $NN$ potential truncated at a given order in the chiral expansion, which includes long-range contributions from pion exchange interactions and short-range contributions parametrized by nucleon contact interactions. The most common scheme used to regularize the ultraviolet (UV) divergences in the LS equation is to introduce a sharp or smooth momentum cutoff regularizing function \cite{epelbaum1a,machleidt1} that suppresses the contributions from the potential matrix elements for momenta larger than a given momentum cutoff scale (multi-pion exchange interactions also involve UV divergent loop integrals which must be consistently regularized and renormalized). In the second step, one has to determine the strengths of the contact interactions, the so called low-energy constants (LEC's), by fitting a set of low-energy scattering data. Once the LEC's are fixed at a given momentum cutoff scale, the LS equation can be solved to evaluate other observables. The $NN$ interactions can be considered properly renormalized when the predicted observables are (approximately) independent of the momentum cutoff scale within the range of validity of ChEFT \cite{machleidt1,lepage}.

The state of the art chiral $NN$ potentials available to date, constructed within the framework of Weinberg's approach to ChEFT, are the next-to-next-to-next-to-leading-order ($N^3LO$) potentials of Epelbaum, Gl\"ockle and Meissner \cite{epelbaum2} and of Entem and Machleidt \cite{entem1}. Both these potentials provide a very accurate description of $NN$ scattering data below laboratory energies $E_{\rm LAB} \sim 300~{\rm MeV}$, with a $\chi^2/{\rm d.o.f}\sim 1$ comparable to that obtained by high-precision phenomenological potentials such as the Nijmegen II \cite{nijmegen} and the Argonne V18 \cite{argonne}, and have been successfully applied in many nuclear structure and reaction calculations. Furthermore, the leading chiral three-nucleon ($3N$) forces at $NNLO$ have been derived \cite{threenf1,threenf2} and applied with fair success in calculations of few-nucleon reactions, light and medium nuclei, and infinite nuclear and neutron matter. Subleading chiral $3N$ forces at $N^3LO$ \cite{threenf3,threenf4,threenf5} and $N^4LO$ \cite{fournf1,fournf2,fournf3} have been recently worked out and are expected to resolve some challenging nuclear structure and reaction problems that remain open. For a review and a comprehensive list of references on the applications of ChEFT to few- and many-nucleon systems, see e.g. \cite{epelbaum1a,machleidt1,epelbaum1b}.

In spite of its remarkable phenomenological success, the theoretical basis of Weinberg's approach has been criticized. In particular, conceptual questions have been raised regarding the formal inconsistency between Weinberg's power counting scheme (based on naive dimensional analysis) and the non-perturbative renormalization of the $NN$ interactions \cite{epelbaum1a,machleidt1}. Such inconsistency arises from the non-renormalizability of ChEFT, which is a consequence of the highly singular nature of the interactions in the chiral expansion of the $NN$ potential at short distances. The iteration of the $NN$ potential truncated at a given order in the chiral expansion (using the LS or the Schr\"odinger equation) generates higher-order UV divergences, and hence strong regularization scale dependencies, which cannot be absorbed by the contact interactions (counterterms) included in the potential at that same order. Thus, upon momentum cutoff regularization, the limit of infinite cutoff cannot be taken while keeping the amplitudes finite and cutoff-independent, i.e. renormalization group invariant. So far, the issue of proper non-perturbative renormalization of chiral $NN$ interactions remains controversial. From one side, in the successful ChEFT calculations based on Weinberg's approach, implemented by using the chiral $N^3LO$ potentials mentioned above, finite cutoffs are typically chosen in the range $\sim 450 - 600~{\rm MeV}$ and nearly cutoff-independent stable results are obtained provided the cutoff is varied only over a narrow window \cite{epelbaum1a,machleidt1,epelbaum1b}. A justification for such a setup with basis on Lepage's view of renormalization in cutoff EFT \cite{lepage} has been presented in the works of Epelbaum and Meissner \cite{epelbaum3} and Epelbaum and Gegelia \cite{gegelia2}. Still within the framework of finite cutoff calculations, it has been shown that the explicit inclusion of $\Delta-{\rm isobar}$ degrees of freedom in ChEFT improves the convergence of the chiral expansion for the $NN$ interactions as compared to the theory with only pion and nucleon degrees of freedom \cite{deltaiso1,deltaiso2,deltaiso3,deltaiso4,deltaiso5,deltaiso6}. From the other side, the criticism to the consistency of Weinberg's approach, and in particular to the narrow range of cutoffs for which it can provide renormalization group invariant results, led to the proposal of alternative renormalization approaches and power counting schemes such as those by Kaplan, Savage an Wise \cite{kaplan1,kaplan2,kaplan3}, van Kolck \cite{bira1}, Nogga et al. \cite{nogga1}, Birse \cite{birse3,birse4,birse5}, Valderrama \cite{valderrama6,valderrama7,valderrama8}, Beane et al. \cite{beane1,beane2} and Long et al. \cite{long1,long2,long3}. In this scenario, methods based on Wilson's renormalization group \cite{wilson1,wilson2,wilson3} have been successfully applied to analyze the scale dependence of chiral $NN$ potentials, both in momentum space \cite{birse3,birse4,birse5,birse1,birse2} and in coordinate space \cite{valderrama1,valderrama2,valderrama3,valderrama4,valderrama5}, providing a better understanding of the interplay between the power counting and the non-perturbative renormalization in ChEFT. Although much progress has been made in this direction, the construction of a consistent framework for the non-perturbative renormalization of $NN$ interactions still remains an open problem \cite{machleidt2}.

An alternative approach to the non-perturbative renormalization of $NN$ interactions that has also been explored is the subtracted kernel method (SKM) \cite{npa99,plb00,hep01,plb05,npa07,ijmpe07,prc11,fb19,efb21,aop2010} in which, instead of using a momentum cutoff regularizing function, the LS equation is regularized by performing multiple subtractions in the kernel at a given energy scale. A closely related subtractive renormalization approach is described in Refs. \cite{yang1,yang2,yang3}, although in that case a sharp momentum cutoff is also introduced. In this work we apply the SKM approach to the chiral $NN$ interactions up to next-to-next-to-leading-order ($NNLO$). We consider the scattering of two nucleons in the $^1 S_0$ channel and analyze the errors in the phase-shifts calculated at each order in Weinberg's power counting scheme (WPC) and in a modified power counting scheme (MPC) based on the promotion of contact interactions which yields a systematic order-by-order power-law improvement. We also show, by explicit numerical calculations, that the SKM procedure is renormalization group invariant under the change of the subtraction scale, provided the renormalized interactions are evolved through a non-relativistic Callan-Symanzik (NRCS) flow equation \cite{CS1,CS2}.

\section{SKM approach for the NN system.}

We start by considering the chiral expansion for the effective $NN$ potential in Weinberg's power counting scheme (WPC) \cite{epelbaum1a,machleidt1}. In a partial-wave relative momentum space basis, the matrix-elements of the $NN$ potential in the $^1S_0$ channel up to $N^3LO$ are given by
\begin{eqnarray}
V^{LO}(p,p') &=& V^{LO}_{1\pi}(p,p') + C_0 \;\\
V^{NLO}(p,p') &=& V^{LO}(p,p') + V^{NLO}_{1\pi}(p,p')+ V^{NLO}_{2\pi}(p,p')+C_2(p^2+p'^2) \;\\
V^{NNLO}(p,p') &=& V^{NLO}(p,p')+ V^{NNLO}_{1\pi}(p,p')+ V^{NNLO}_{2\pi}(p,p')\;\\
V^{N^3LO}(p,p') &=& V^{NNLO}(p,p')+ V^{N^3LO}_{1\pi}(p,p') + V^{N^3LO}_{2\pi}(p,p')\nonumber\\
                &+& V^{N^3LO}_{3\pi}(p,p')+ C'_4~p^2 p'^2 + C_4(p^4 + p'^4) \; ,
\end{eqnarray}
\noindent
where the coefficients $C_i$ stand for the strengths of the contact interactions and an obvious notation is used for the pion-exchange interactions.

The matrix-elements of the leading-order ($LO$) OPE potential in the $^1S_0$ channel are given by
\begin{equation}
V^{LO}_{1\pi}(p,p')=\frac{g_a^2}{32\pi f_\pi^2}
\left(2-\int^1_{-1}dx \frac{m_\pi^2}{p^2+p'^2-2 p p'x+m_\pi^2}\right) \; ,
\label{VOPE1S0}
\end{equation}
where $g_a$, $f_\pi$ and $m_\pi$ denote, respectively, the axial coupling constant, the pion weak-decay constant and the pion mass. The higher-order OPE terms include corrections from pion loops and counter term insertions, which only contribute to the renormalization of coupling constants and masses. In this work, we use $g_a=1.25$, $f_\pi =93~{\rm MeV}$ and $m_\pi=138~{\rm MeV}$. The two-pion-exchange (TPE) potential is taken from Ref. \cite{epelbaum2}.

Consider the formal LS equation for the $T$-matrix of a two-nucleon system, which can be written in operator form as
\begin{eqnarray}
T(E) &=& V + V~G_{0}^{+}(E)~T(E) \; ,
\label{LS}
\end{eqnarray}
where $E$ is the energy of the two-nucleon system in the center-of-mass frame, $V$ is the effective $NN$ potential and $G_{0}^{+}(E)= (E - H_{0} + i \epsilon)^{-1}$ is the free Green's function for the two-nucleon system with outgoing-wave boundary conditions, given in terms of the free hamiltonian $H_0$. Both pion-exchange and contact interaction terms can lead to UV divergences when the effective $NN$ potential $V$ at a given order in the chiral expansion is iterated in the LS equation, requiring a regularization and renormalization procedure in order to obtain well-defined finite solutions.

In the standard cutoff renormalization scheme the formal LS equation, Eq. (\ref{LS}), is regularized by multiplying the effective $NN$ potential $V$ with a momentum cutoff regularizing function. The common choice is an exponential $f(p)=\exp[-(p/\Lambda)^{2r})]$ (with $r=1, 2, \ldots$), where $\Lambda$ is a cutoff parameter, such that
\begin{equation}
 V(p, p') \rightarrow  V_{\Lambda}(p, p') \equiv \exp[-(p/\Lambda)^{2r})]~ V(p, p')~ \exp[-(p'/\Lambda)^{2r})] \; .
 \label{CR}
\end{equation}

In the SKM approach, a regularized and renormalized LS equation for the $T$-matrix is computed through an iterative procedure which involves recursive multiple subtractions performed in kernel at a given energy scale. For a general number of subtractions $n$, we define a $n$-fold subtracted LS equation given in operator form by
\begin{equation}
T^{(n)}_{\mu}(E) = V^{(n)}_{\mu}(E) + V^{(n)}_{\mu}(E)~G_{n}^{+}(E;-\mu^2)~T^{(n)}_{\mu}(E) \; ,
\label{LSn}
\end{equation}
where $\mu$ is the subtraction scale, $V^{(n)}_{\mu}(E)$ is called ``driving term" and $G_{n}^{+}(E;-\mu^2)$ is the $n$-fold subtracted Green's function, defined by
\begin{equation}
G_{n}^{+}(E;-\mu^2) \equiv \left(\frac{\mu^2+E}{\mu^2+H_0} \right)^{n}~G_{0}^{+}(E)= F_{n}(E;-\mu^2)~G_{0}^{+}(E)  \; .
\label{Gn}
\end{equation}
\noindent
Note that we choose a negative energy subtraction point $-\mu^2$, such that the free Green's function $G_{0}^{+}(-\mu^2)$ is real.

The $n$-fold subtracted LS equation, Eq. (\ref{LSn}), has the same operator structure as the formal LS equation, Eq. (\ref{LS}), but with the effective $NN$ potential $V$ replaced by the driving term $V^{(n)}_{\mu}(E)$ and the free Green's function $G_{0}^{+}(E)$ replaced by the $n$-fold subtracted Green's function $G_{n}^{+}(E;-\mu^2)$. Moreover, regularization is achieved not by using a momentum cutoff regularizing function, but instead through the form factor $F_{n}(E;-\mu^2)$ introduced by the $n$-fold subtracted Green's function, which is built up as part of the iterative procedure. The driving term encodes the physical information apparently lost due to the removal of the propagation through intermediate states at the subtraction scale $\mu$. Thus, once the driving term is determined (by fixing the strengths of the contact interactions at the subtraction scale $\mu$) the subtracted LS equation provides a renormalized finite solution for the $T$-matrix at any given energy $E$.

The driving term $V^{(n)}_{\mu}(E)$ is recursively constructed through an iterative procedure starting from $V^{(1)}_{\mu}(E=-\mu^2)\equiv T^{(1)}_{\mu}(E=-\mu^2)$, which is replaced by an {\it ansatz} for the $T$-matrix at the subtraction scale, $T(-\mu^2)$. The recursion formula (with $m=1,...,n$) is given by

\begin{equation}
V^{(m)}_{\mu}(E) = {\bar V}^{(m)}_{\mu}(E) + V^{(m)}_{\mu,~{\rm sing}}  \; ,
\label{Vm}
\end{equation}
\noindent
where
\begin{equation}
{\bar V}^{(m)}_{\mu}(E) = \left[1-(-\mu^2-E)^{m-1}V^{(m-1)}_{\mu}(E) G_{0}^{+}(-\mu^2)^m \right]^{-1}V^{(m-1)}_{\mu}(E) \; ,
\label{Vbarm}
\end{equation}
\noindent
and the term $V^{(m)}_{\mu,~{\rm sing}}(E)$ contains the higher-order singular interactions that generate divergent integrals which can be regularized by performing $m$ subtractions. One should note from Eq. (\ref{Vm}) the that the driving term $V^{(m)}_{\mu}(E)$ at each iteration is derived in two steps. First, we calculate ${\bar V}^{(m)}_{\mu}(E)$ from $V^{(m-1)}_{\mu}(E)$, solving an integral equation obtained by manipulating Eq. (\ref{Vbarm}):
\begin{equation}
{\bar V}^{(m)}_{\mu}(E) =V^{(m-1)}_{\mu}(E)+(-\mu^2-E)^{m-1}~V^{(m-1)}_{\mu}(E)~G_{0}^{+}(-\mu^2)^m ~{\bar V}^{(m)}_{\mu}(E)  \; .
\label{Vbarmint}
\end{equation}
\noindent
Then, we introduce the corresponding higher-order singular interactions in the driving term by adding $V^{(m)}_{\mu,~{\rm sing}}(E)$.

For convenience, in this work we implement the SKM procedure using the $K$-matrix instead of the $T$-matrix. The LS equation for the $K$-matrix in the $^1S_0$ channel with $n$ subtractions is given by (here and in what follows we use units such that $\hbar=c=M=1$, where $M$ is the nucleon mass)
\begin{eqnarray}
K^{(n)}_{\mu}(p,p';k^2) &=& V^{(n)}_{\mu}(p,p';k^2) + \frac 2 \pi \mathcal{P} \int_0^\infty dq~q^2
\left(\frac{\mu^2+k^2}{\mu^2+q^2}\right)^n \nonumber\\&\times& \frac{V^{(n)}_{\mu}(p,q;k^2)}{k^2-q^2}K^{(n)}_{\mu}(q,p';k^2) \; ,
\label{subkn}
\end{eqnarray}
\noindent
where $k=\sqrt{E}$ is the on-shell momentum in the center-of-mass frame and ${\cal P}$ denotes the principal value. Note that the $n$-fold subtracted Green's function introduces an energy- and $\mu$-dependent form factor proportional to $q^{-2n}$ in the kernel of the subtracted LS equation which regularizes UV power divergences up to order ${\cal O}(q^{2n-1})$, thus effectively acting like a smooth momentum cutoff regularizing function.

At each order in the chiral expansion we perform the minimum number of subtractions $n$ necessary to render a finite solution for the subtracted $K$-matrix. The corresponding driving term $V^{(n)}_{\mu}$ is computed through the iterative procedure described above, starting from the {\it ansatz} for the $LO$ driving term $V^{(1)}_{\mu}$ given by
\begin{equation}
V^{(1)}_{\mu}(p,p';-\mu^2) = V^{LO}_{1\pi}(p,p') + C_0(\mu) \; .
\label{v1}
\end{equation}
\noindent
One should note that only one subtraction is enough to get a finite result for the $K$-matrix at $LO$, since when iterated in the LS equation the $LO$ potential generates UV power divergences of order ${\cal O}({q^1})$.

At next-to-leading-order ($NLO$) we have to construct a 3-fold subtracted kernel LS equation in order to get a finite result for the $K$-matrix, since when iterated in the LS equation the $NLO$ potential generates UV power divergences up to order ${\cal O}({q^5})$, even thought it contains only terms up to order ${\cal O}({q^2})$. To obtain the $NLO$ driving term, $V^{(3)}_{\mu}$, we first calculate ${\bar V}^{(2)}_{\mu}$ from $V^{(1)}_{\mu}$,
\begin{eqnarray}
{\bar V}^{(2)}_{\mu}(p,p';k^2) &=& V^{(1)}_{\mu}(p,p';k^2)- \frac 2 \pi \int_0^\infty dq~q^2 ~ V^{(1)}_{\mu}(p,q;k^2)
\nonumber\\&\times&\frac{(\mu^2+k^2)^{1}}{(\mu^2+q^2)^2}{\bar V}^{(2)}_{\mu}(q,p';k^2) \; .
\label{v2}
\end{eqnarray}
\noindent
Then, we calculate ${\bar V}^{(3)}_{\mu}$ from ${\bar V}^{(2)}_{\mu}$,
\begin{eqnarray}
{\bar V}^{(3)}_{\mu}(p,p';k^2) &=& {\bar V}^{(2)}_{\mu}(p,p';k^2)- \frac 2 \pi \int_0^\infty dq~q^2 ~{\bar V}^{(2)}_{\mu}(p,q;k^2)
\nonumber\\&\times&\frac{(\mu^2+k^2)^{2}}{(\mu^2+q^2)^3}{\bar V}^{(3)}_{\mu}(q,p';k^2) \; ,
\label{v3}
\end{eqnarray}
\noindent
and add the $NLO$ interactions:
\begin{eqnarray}
V^{(3)}_{\mu}(p,p';k^2)&=& {\bar V}^{(3)}_{\mu}(p,p';k^2) + V^{NLO}_{2\pi}(p,p')+ C_2(\mu)~(p^2+p'^2) \; .
\label{v3nlo}
\end{eqnarray}

At $NNLO$ we have to construct a 4-fold subtracted kernel LS equation in order to get a finite result for the $K$-matrix. Even thought the $NNLO$ potential contains only terms up to order ${\cal O}({q^3})$, when iterated in the LS equation it generates UV power divergences up to order ${\cal O}({q^7})$. To obtain the $NNLO$ driving term, $V^{(4)}_{\mu}$, we first calculate ${\bar V}^{(4)}_{\mu}$ from $V^{(3)}_{\mu}$,
\begin{eqnarray}
{\bar V}^{(4)}_{\mu}(p,p';k^2) &=& V^{(3)}_{\mu}(p,p';k^2)- \frac 2 \pi \int_0^\infty dq~q^2 ~ V^{(3)}_{\mu}(p,q;k^2)
\nonumber\\&\times&\frac{(\mu^2+k^2)^{3}}{(\mu^2+q^2)^{4}}{\bar V}^{(4)}_{\mu}(q,p';k^2) \; ,
\label{v4}
\end{eqnarray}
\noindent
and then we add the $NNLO$ interactions:
\begin{equation}
V^{(4)}_{\mu}(p,p';k^2)= {\bar V}^{(4)}_{\mu}(p,p';k^2)+V^{NNLO}_{2\pi}(p,p') \; .
\label{v4nnlo}
\end{equation}

The renormalized strengths $C_i(\mu)$ of the contact interactions included in the driving term $V^{(n)}_{\mu}$ at each order in the chiral expansion are fixed at the subtraction scale $\mu$ by fitting data for low-energy scattering observables, thus encoding the input physical information. Instead of the usual matching of scattering data at discrete values of the on-shell momentum $k$, we follow the procedure described by Steele and Furnstahl \cite{steele1,steele2}, which is numerically much more robust. Here, we use as ``data" the values of the inverse on-shell $K$-matrix evaluated from the solution of the LS equation with the Nijmegen-II potential \cite{nijmegen} for a spread of very small momenta $k~(\le 0.1~{\rm fm^{-1}})$ and fit the difference between such data and the inverse on-shell $K$-matrix evaluated from the solution of the $n$-fold subtracted LS equation (Eq.~(\ref{subkn})) with the driving term $V^{(n)}_{\mu}$ to an interpolating polynomial in $k^2/\mu^2$ to highest possible degree,
\begin{eqnarray}
\Delta(1/K)&=&1/K_{{\rm NIJ}}(k,k;k^2)-1/K^{(n)}_{\mu}(k,k;k^2)\nonumber\\
&=& A_0 + A_2\;\frac{k^2}{\mu^2} + A_4\;\frac{k^4}{\mu^4} + \ldots \;.
\end{eqnarray}
\noindent
The coefficients $A_i$ are then minimized with respect to the variations in the renormalized strengths $C_i(\mu)$. One should note that the procedure outlined above is equivalent to fix the renormalized strengths $C_i(\mu)$ by fitting the experimental values of the parameters of the effective range expansion (ERE) \cite{bethe} to a given order in $k^2$, since at very low energies the phase-shifts provided by the ERE agree very well with those obtained from the Nijmegen-II potential, i.e.
\begin{eqnarray}
-1/K_{{\rm NIJ}}(k,k;k^2)&=& k~\cot\delta_{{\rm NIJ}}(k)\simeq k~\cot\delta_{{\rm ERE}}(k)\nonumber\\
&=&-\frac{1}{a}+\frac{1}{2}r_e k^2 + v_2 k^4 + v_3 k^6 + \ldots,
\end{eqnarray}
\noindent
where $a$ is the scattering length, $r_e$ is the effective range and $v_i$ are the shape parameters.

\section{Power counting for the NN system in the $^1 S_0$ channel.}

Once the renormalized strengths $C_i(\mu)$ are fixed at the subtraction scale $\mu$, and so the driving term $V^{(n)}_{\mu}$ is determined, we can calculate the $NN$ scattering observables for any given energy from the numerical solution of the $n$-fold subtracted LS equation for the $K$-matrix (Eq.~(\ref{subkn})). In log-log plots for the relative errors in the observables (``Lepage plots"), we expect to obtain straight lines with slopes given by the dominant power of $k^2/\mu^2$ in the errors. A systematic power-law improvement in the predictions of the observables is expected as more contact interactions are included in the driving term $V^{(n)}_{\mu}$ and the corresponding renormalized strengths $C_i(\mu)$ are fixed. The impact of the included long-range pion-exchange interactions on the scaling of the errors is, however, much less transparent and so must be verified through the explicit numerical calculation of the observables.
\begin{figure*}[t]
\centerline{\includegraphics[scale=0.75]{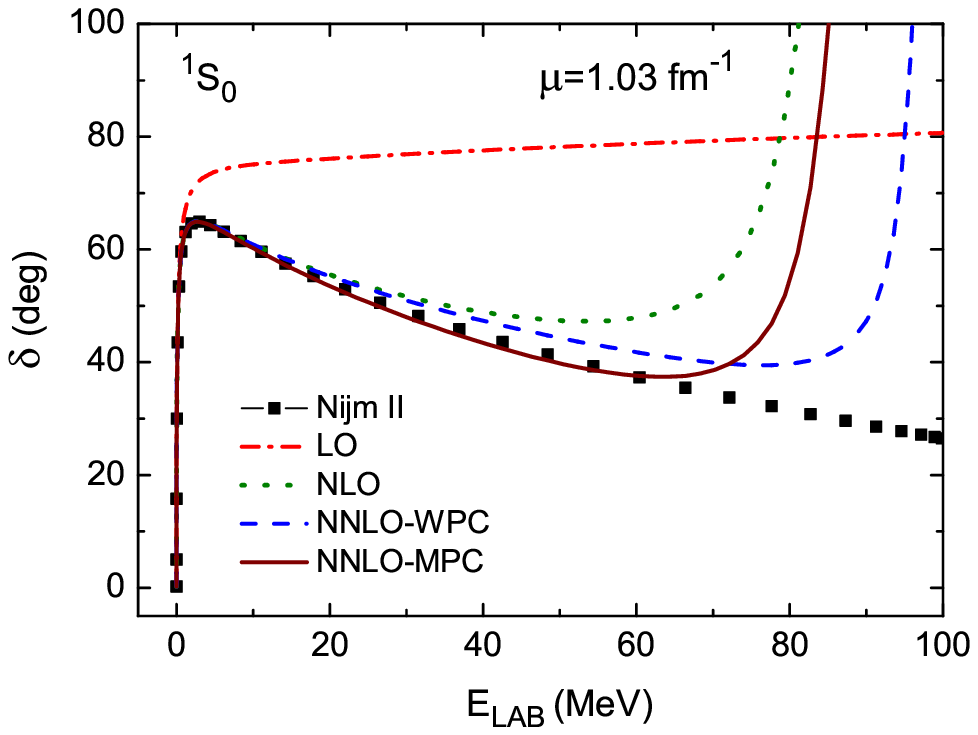}\hspace*{0.6cm}\includegraphics[scale=0.75]{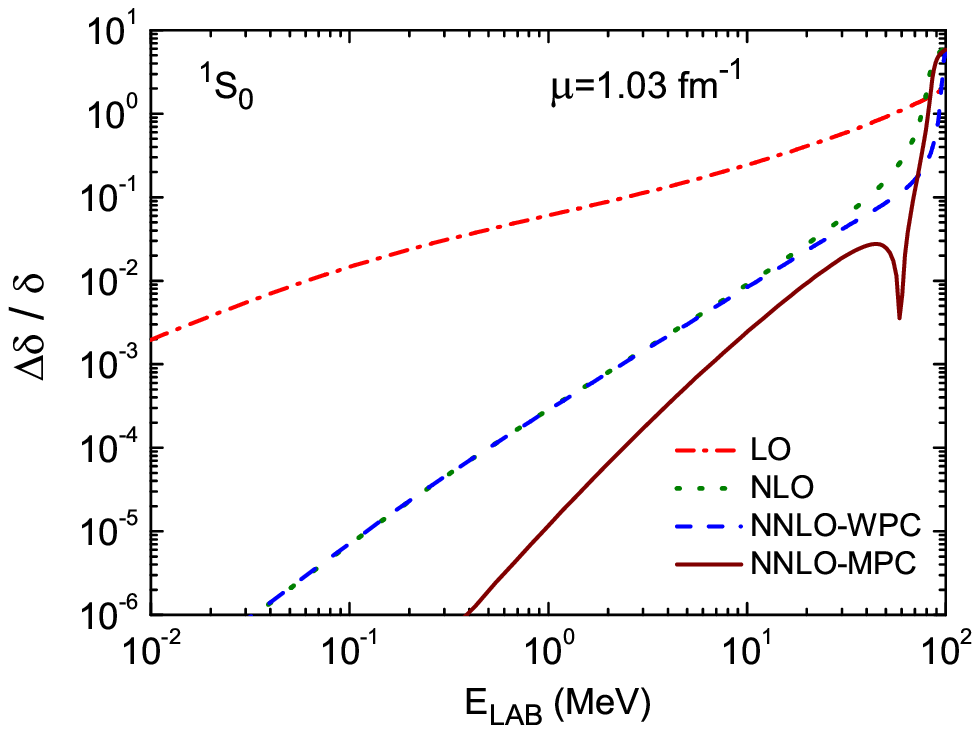}}
\caption{(Color on-line) Phase-shifts in the $^1 S_0$ channel calculated from the solution of the subtracted LS equation for the $K$-matrix at $LO$, $NLO$ and $NNLO$ both in Weinberg's power counting scheme (WPC) and in the modified power counting scheme (MPC) with the renormalized strengths $C_i(\mu)$ fixed at $\mu=1.03~{\rm fm^{-1}}$ (left) and the corresponding relative errors with respect to the Nijmegen-II phase-shifts (right).}
\label{fig1}
\end{figure*}

In the left panel of Fig.~\ref{fig1} we show the phase-shifts in the $^1 S_0$ channel as a function of the laboratory energy ($E_{\rm LAB}$) calculated from the numerical solution of the $n$-fold subtracted LS equation for the $K$-matrix at $LO$ ($n=1$), $NLO$ ($n=3$) and $NNLO$ ($n=4$), with the renormalized strengths $C_i(\mu)$ of the contact interactions included at each order fixed at the subtraction scale $\mu=1.03~{\rm fm^{-1}}$. In the right panel we show the log-log plots for the corresponding relative errors with respect to the Nijmegen-II phase-shifts. For the phase-shifts calculated at $LO$, where the driving term consists of the OPE interaction plus the non-derivative contact interaction, we obtain the well-known strong deviation from the Nijmegen-II results, with the relative errors scaling like ${\cal O}(k^2/\mu^2)$ at low-energies. For the phase-shifts calculated at $NLO$, where a TPE interaction term and the second-order derivative contact interaction are included, we obtain the expected power-law improvement, with the relative errors scaling like ${\cal O}(k^4/\mu^4)$ at low-energies. At $NNLO$ in the WPC scheme, where only a TPE interaction term is included but no new contact interaction, we obtain a better overall agreement with the Nijmegen-II phase-shifts but there is no power-law improvement.

We then consider a modified power counting scheme (MPC) in which the fourth-order $^1 S_0$ channel derivative contact interaction $C'_4(\mu)~p^2 p'^2$ is promoted from $N^3LO$ to $NNLO$ and, therefore, added to the driving term $V^{(4)}_{\mu}$ given by Eq.~(\ref{v4nnlo}). As one can see from the error plots shown in the right panel of Fig.~\ref{fig1}, at $NNLO$ in the MPC scheme the relative errors in the phase-shifts scale like ${\cal O}(k^6/\mu^6)$ at low-energies and so we obtain a systematic order-by-order power-law improvement.

Modifications of the WPC scheme based on the promotion of contact interactions have been considered in several works \cite{nogga1,birse3,birse4,birse5,valderrama6,valderrama7,valderrama8,long1,long2,long3}. In the power counting scheme proposed by Nogga, Timmermans and van Kolck (NTvK) \cite{nogga1}, which is also used in the nuclear matter calculations described in Ref.~\cite{machleidt3}, higher-order contact interactions are promoted to $LO$ in low angular momentum partial-wave channels where the OPE tensor interaction is singular and attractive (namely, in the $^3 P_0$, $^3 P_2~-~^3 F_2$ and $^3 D_2$ channels), such that reasonably cutoff independent results are obtained in $LO$ calculations for cutoffs varying in the range $2 \dots 20~{\rm fm^{-1}}$. A criticism of the NTvK scheme was made in Ref.~\cite{epelbaum3} where, based on an analysis of several $NN$ scattering observables calculated at $LO$ both in the NTvK and the WPC scheme, it has been argued that the use of larger cutoffs and the modifications of the WPC scheme as proposed in Ref.~\cite{nogga1} do not improve the results and can even lead to discrepancies in certain partial-wave channels.

Our MPC scheme resembles that implemented by Valderrama \cite{valderrama6,valderrama7,valderrama8} within the framework of a perturbative treatment of the chiral TPE interactions at $NLO$ and $NNLO$, which can be regarded as an extension of the NTvK scheme to subleading orders. In Valderrama's approach, the $LO$ phase-shifts are calculated non-perturbatively from the solution of the Schr{\"o}dinger equation in coordinate space with the $LO$ potential (OPE interaction plus the non-derivative contact interactions) iterated to all orders. The contributions to the phase-shifts from the subleading order terms of the potential are included perturbatively within the distorted-wave Born approximation ($DWBA$), resulting in a set of power counting rules (based on the requirement of renormalizability of the scattering amplitude) which are in agreement with the modifications of the WPC scheme determined in the renormalization group analysis made by Birse \cite{birse3,birse4,birse5} (with some minor differences). A perturbative treatment of the subleading order interactions is also considered in Refs.~\cite{beane1,beane2,long1,long2,long3,birse6,birse7,birse8,birse9}. Here, we consider the application of the SKM approach to the chiral $NN$ interactions within a non-perturbative renormalization framework and so we calculate the phase-shifts by iterating the full $NN$ potential (LO plus subleading order terms) to all orders in the subtracted LS equation. One should also note that in Valderrama's approach the power counting scheme is such that it requires the promotion of the $^1 S_0$ channel contact interaction $C_4~(p^4 + p'^4)$ from $N^3LO$ to $NLO$, while in our MPC scheme we promote the $^1 S_0$ channel contact interaction $C'_4~p^2 p'^2$ from $N^3LO$ to $NNLO$.

In principle, we could promote either of the two $^1 S_0$ channel fourth-order derivative contact interactions available at $N^3LO$ in the WPC scheme. We choose to promote the $C'_4(\mu)~p^2 p'^2$ contact interaction because when iterated in the standard LS equation it generates additional UV power divergences only up to order ${\cal O}({q^6})$ (due to the presence of the $NNLO$ TPE interaction term), which can be regularized by performing the same four subtractions as required for the $NNLO$ calculation in the WPC scheme. The iteration of the $C_4(\mu)~(p^4 + p'^4)$ contact interaction, on the other hand, generates additional UV power divergences up to order ${\cal O}({q^9})$, and so would require at least five subtractions. Moreover, with the $C_4(\mu)~(p^4 + p'^4)$ contact interaction left to be included at $N^3LO$, there will be no need to promote a higher-order contact interaction to obtain the power-law improvement at this order. In this way, using such a prescription we would obtain in our MPC scheme the same number of $^1 S_0$ channel contact interactions at $N^3LO$ as predicted in the WPC scheme, which is the one employed in the construction of the state of the art chiral $N^3LO$ chiral potentials described in Refs. \cite{epelbaum2} and \cite{entem1}.
\begin{figure*}[t]
\centerline{\includegraphics[scale=0.75]{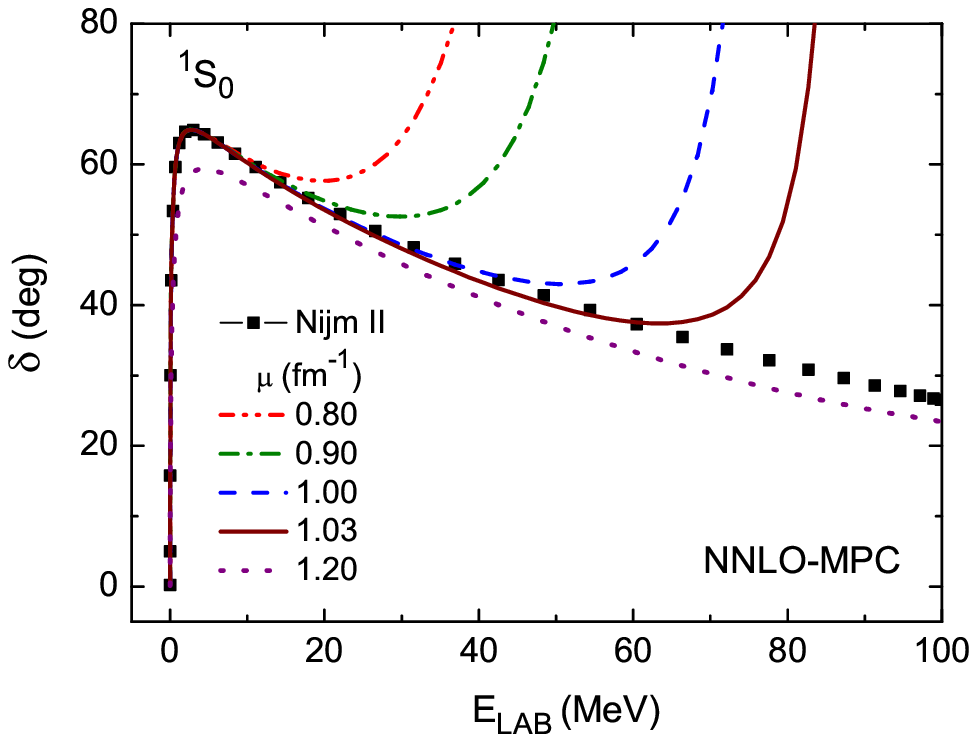}\hspace*{0.6cm}\includegraphics[scale=0.75]{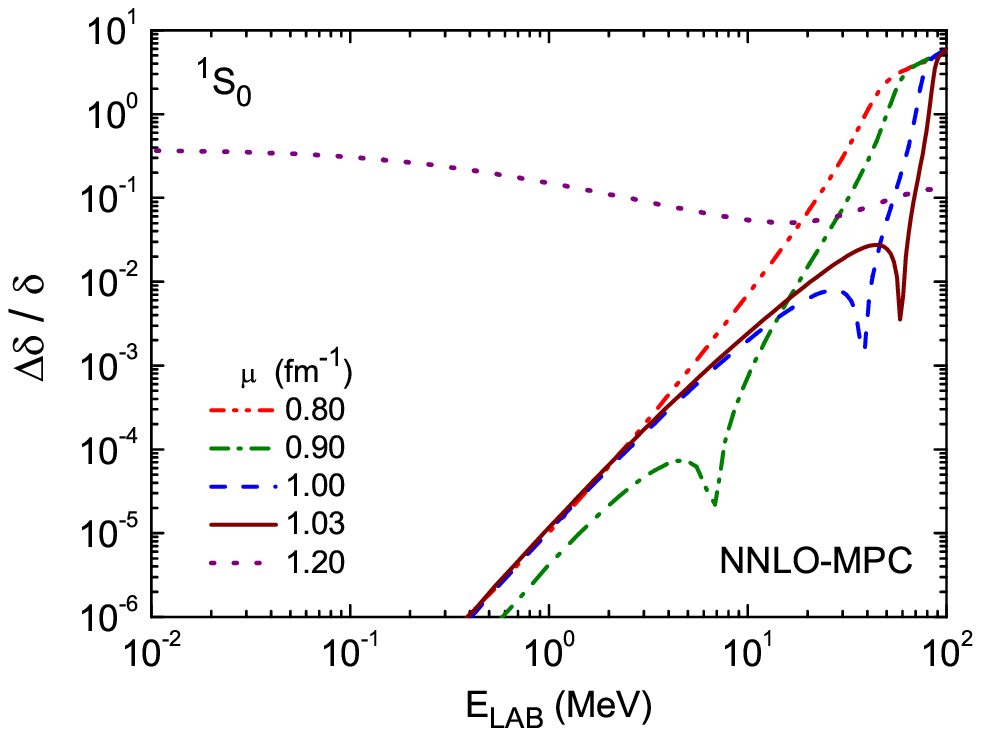}}
\caption{(Color on-line) Phase-shifts in the $^1 S_0$ channel calculated from the solution of the subtracted LS equation for the $K$-matrix at $NNLO$ in the modified power counting scheme (MPC) for several values of the subtraction scale $\mu$ (left) and the corresponding relative errors with respect to the Nijmegen-II phase-shifts (right).}
\label{fig2}
\end{figure*}

In the left panel of Fig.~\ref{fig2} we show the phase-shifts in the $^1S_0$ channel as a function of $E_{\rm LAB}$ calculated from the solution of the subtracted LS equation for the $K$-matrix at $NNLO$ in the MPC scheme, with the renormalized strengths $C_i(\mu)$ ($i=0,~2,~4$) fixed at several values of the subtraction scale $\mu$. In the right panel we show the log-log plots for the corresponding relative errors with respect to the Nijmegen-II phase-shifts. As one can see, the range of energies for which the SKM procedure provides a good description of the phase-shifts increases with $\mu$. For $\mu = 1.03~{\rm fm^{-1}}$, the SKM results are in good agreement with the Nijmegen-II phase-shifts up to $E_{\rm LAB} \sim 60~{\rm MeV}$.

It is important to observe that by performing an overall best-fit (instead of using the ``data" for on-shell momenta $k\le 0.1~{\rm fm^{-1}}$) a fairly good description of the phase-shifts can be obtained up to $E_{\rm LAB}\sim 200~{\rm MeV}$, but that would prevent us from making a proper error scaling analysis for predicted phase-shifts. Furthermore, as a consequence of the Wigner causality bound \cite{wigner1} we find that by taking a subtraction scale $\mu$ larger than $\sim 1.1~{\rm fm^{-1}}$ it becomes impossible to fit the inverse on-shell $K$-matrix evaluated from the Nijmegen-II potential to within an acceptable accuracy while keeping the $NNLO$ driving term $V^{(4)}_{\mu}$ hermitian (i.e., with the renormalized strengths $C_i(\mu)$ taking only real values), such that the unitarity of the scattering amplitude is preserved. This is illustrated by the results obtained for $\mu = 1.2~{\rm fm^{-1}}$, which cannot reproduce well the Nijmegen-II phase-shifts even at very low energies, and so clearly indicate the breakdown of the fitting procedure. Several works have discussed the implications of the Wigner causality bound \cite{deltaiso5,valderrama2,valderrama4,cohen1,cohen2,cohen3,cohen4,cohen5,gegelia1} and the so called low-energy theorems \cite{gegelia2,cohen6,cohen7} to the renormalization problem, both in the context of pionless EFT and ChEFT. The results described in the works cited above give strong support to the prescription advocated by Lepage \cite{lepage} that in order to get a consistent cutoff EFT for the $NN$ interactions the cutoff should 
not be taken much larger than the relevant hard scale in the theory, e.g. the pion mass $m_{\pi}\sim 140 \; {\rm MeV}$ in the case of pionless EFT and the chiral symmetry breaking scale $\Lambda_{\chi} \sim 1 \; {\rm GeV}$ in the case of ChEFT. The most efficient choice is to take the cutoff of the order of the relevant hard scale \cite{epelbaum1a,lepage,epelbaum3,gegelia2}.

EFT's are essentially constructed as systematic low-energy expansions in powers of the ratio ($Q/ \Lambda_{\rm hard}$), where $Q$ and $\Lambda_{\rm hard}$ stand respectively for the relevant soft (low-momentum) and hard (high-momentum) scales in the theory, which provide a valid description of phenomena at momentum scales below $\Lambda_{\rm hard}$. In the renormalization of a cutoff EFT for the $NN$ interactions, an UV regularizing momentum cutoff scale $\Lambda$ is introduced in order to remove high-momentum degrees of freedom which can probe the unknown short-distance dynamics, thus playing the role of a resolution scale. The contributions from low-momentum/long-distance degrees of freedom are included explicitly in the calculation of observables through interactions known from the underlying theory (e.g. pion-exchange interactions). The contributions from the excluded high-momentum/short-distance degrees of freedom are included implicitly through parametrized cutoff-dependent contact interactions (counterterms) whose coefficients are fixed by fitting low-energy scattering data, systematically removing cutoff dependence in the observables. Lepage's prescription is based on the view that by taking the momentum cutoff scale $\Lambda$ much larger than $\Lambda_{\rm hard}$ (i.e., beyond the range of validity of the EFT) one is certainly incorporating contributions from incorrect high-momentum/short-distance dynamics which can lead to pathologies, such as the violation of the Wigner causality bound, and even the breakdown of the EFT systematics \cite{gegelia2,machleidt2}.

The Wigner causality bound is a general result, originally derived by Wigner \cite{wigner1} assuming only the physical principles of causality and unitarity, which shows that for a hermitian potential that vanishes beyond some range $R$ there is a lower bound on the derivative of the phase-shifts $\delta(k)$ with respect to the on-shell momentum $k$,
\begin{equation}
\frac{d\delta(k)}{dk} \geq - R + \frac{1}{2 k}\sin(2 \delta(k) + 2 k R) \; .
\end{equation}
\noindent
An alternative derivation was presented later by Phillips and Cohen \cite{cohen1}, who have shown that Wigner's bound yields a constraint on the effective range $r_e$, given by
\begin{equation}
r_e > 2 \left(R -\frac{R^2}{a} + \frac{R^3}{3a^2} \right) \; ,
\label{wbound}
\end{equation}
\noindent
where $a$ is the scattering length. Moreover, it was shown that this constraint still applies when the potential is not identically zero beyond the range $R$, but fall off sufficiently fast for the wave-function to approach its asymptotic solution rapidly enough.

Eq.~~(\ref{wbound}) shows that in the limit of a zero-range interaction ($R \rightarrow 0$) the effective range $r_e$ cannot be positive. It also shows that there is a minimum range $R_{\rm min}$ for which the interaction can reproduce a given scattering length $a$ and effective range $r_e$. In the case of $NN$ scattering in the $^1 S_0$ channel, for which the experimental values of the scattering length and the effective range are respectively given by $a=-23.7~{\rm fm}$ and $r_e = 2.77~{\rm fm}$, this minimum range is $R_{\rm min}=1.3~{\rm fm}$. Furthermore, Scaldeferri et. al. \cite{cohen2} have shown that this constraint is a general feature of contact interactions which still holds even when long-range pion-exchange interactions are explicitly included. In particular, it was shown that an absolute lower bound $R_{\rm min}=1.1~{\rm fm}$ is obtained for the $NN$ scattering in the $^1 S_0$ channel when both the OPE interaction and an arbitrary short-range contact interaction are included.

By identifying the range $R$ with the momentum cutoff scale $\Lambda$ introduced in the renormalization of cutoff EFT for the $NN$ interactions (i.e., $R \sim 1/\Lambda$), Wigner's bound implies both for pionless EFT and ChEFT that in order to obtain a description of phase-shifts which agrees with experimental low-energy $NN$ scattering data, the momentum cutoff scale cannot be removed by taking the limit $\Lambda \rightarrow \infty$ while maintaining causality and unitarity, as required for a consistent EFT framework \cite{deltaiso5,cohen3,cohen4,cohen5}. Indeed, by investigating the renormalizability of chiral $NN$ interactions in the $^1 S_0$ channel at $NLO$ and $NNLO$ in the WPC scheme within the framework of cutoff renormalization, Entem et.al. \cite{deltaiso5} have found that there is a maximum value $\Lambda_{\rm max}$ for the momentum cutoff scale $\Lambda$ above which one cannot fix the strengths of the contact interactions $C_0(\Lambda)$ and $C_2(\Lambda)$ by fitting the experimental values of both the scattering length $a$ and the effective range $r_e$ while keeping the renormalized potential hermitian. For $\Lambda > \Lambda_{\rm max} \sim 1.8~\rm{fm}^{-1}$ in the case of pionless EFT and $\Lambda > \Lambda_{\rm max} \sim 2.5~\rm{fm}^{-1}$ in the case of ChEFT, the strengths $C_0(\Lambda)$ and $C_2(\Lambda)$ diverge before taking complex values. Thus, in both cases the renormalized potential must become non-Hermitian in order to match the renormalization conditions, such that the unitarity of the scattering amplitude is violated (nevertheless, the corresponding phase-shifts remain real). These results are similar to those we have obtained in our calculations for chiral $NN$ interactions in the $^1 S_0$ channel at $NLO$ and $NNLO$ within the framework of the SKM approach. The rather smaller value for the maximum subtraction scale $\mu_{\rm max} \sim 1.1~\rm{fm}^{-1}$ we have found in the SKM approach as compared to $\Lambda_{\rm max}\sim 2.5~\rm{fm}^{-1}$ found in the cutoff renormalization scheme is a consequence of the highly non-trivial energy, momentum and subtraction scale dependence of the scattering amplitude $K^{(n)}_{\mu}$ obtained from the solution of the $n$-fold subtracted LS equation, Eq. (\ref{subkn}), which yields a very different running of the renormalized strengths of the contact interactions $C_i(\mu)$ with the subtraction scale $\mu$ as determined by the renormalization conditions.

It is important to emphasize that both in the cutoff renormalization scheme and the SKM approach, Wigner's bound arises from the non-linear structure of the renormalization conditions relating the strengths of the contact interactions $C_i$ to the ERE parameters ($a$, $r_e$, $v_2$, etc) used as physical input \cite{cohen3,cohen4,cohen5}. Therefore, the constraint on the maximum value of the regularizing scale for which one can fit the experimental values of the ERE parameters (beyond zeroth order) with a hermitian potential (i.e., with the strengths of the contact interactions $C_i$ taking only real values) applies regardless of the particular procedure used to implement the fitting.

\section{Renormalization group invariance in the SKM approach.}

As pointed before, the multiple subtractions performed in the SKM procedure introduce a form factor in the kernel of the LS equation which acts like a regularizing function, such that the subtraction scale $\mu$ ends up playing a role similar to that of a smooth momentum cutoff scale. But the subtraction scale $\mu$ is arbitrary, and so the scattering observables calculated from the solution of the subtracted LS equation for the scattering amplitude should not depend on its particular choice. By requiring the fully off-shell $K$-matrix with $n$ subtractions to be invariant under the change of the subtraction scale $\mu$, a renormalization group (RG) equation can be derived for the driving term $V^{(n)}_{\mu}(E)$ in the form of a non-relativistic Callan-Symanzik (NRCS) flow equation \cite{plb00}, which is given in operator form by
\begin{equation}
\frac{\partial V^{(n)}_{\mu}(E)}{\partial\mu^2} = -V^{(n)}_{\mu}(E)~\frac{\partial G_{n}^{+}(E;-\mu^2)}{\partial\mu^2}~V^{(n)}_{\mu}(E) \; ,
\label{CSEn}
\end{equation}
\noindent
with the boundary condition $V^{(n)}_{\mu}|_{\mu \rightarrow {\bar \mu}}= V^{(n)}_{\bar \mu}$ imposed at some reference subtraction scale $\bar \mu$ where the renormalized strengths of the contact interactions $C_i(\mu)$ are fixed to fit low-energy observables used as physical input.

The NRCS flow equation for the driving term $V^{(n)}_{\mu}$ is similar to the RG equation for the $NN$ potential derived by Birse in the context of cutoff EFT \cite{birse1,birse2,birse5}, which is also obtained from the invariance of the off-shell $K$-matrix. Another similar RG equation is that derived by Bogner et al. in the $V_{\rm low-k}$ approach \cite{bogner1,bogner2,bogner3}, which is based on the invariance of the half-on-shell $T$-matrix such that it also involves the iteration of the scattering amplitude.
\begin{figure*}[t]
\centerline{
\includegraphics[scale=0.38]{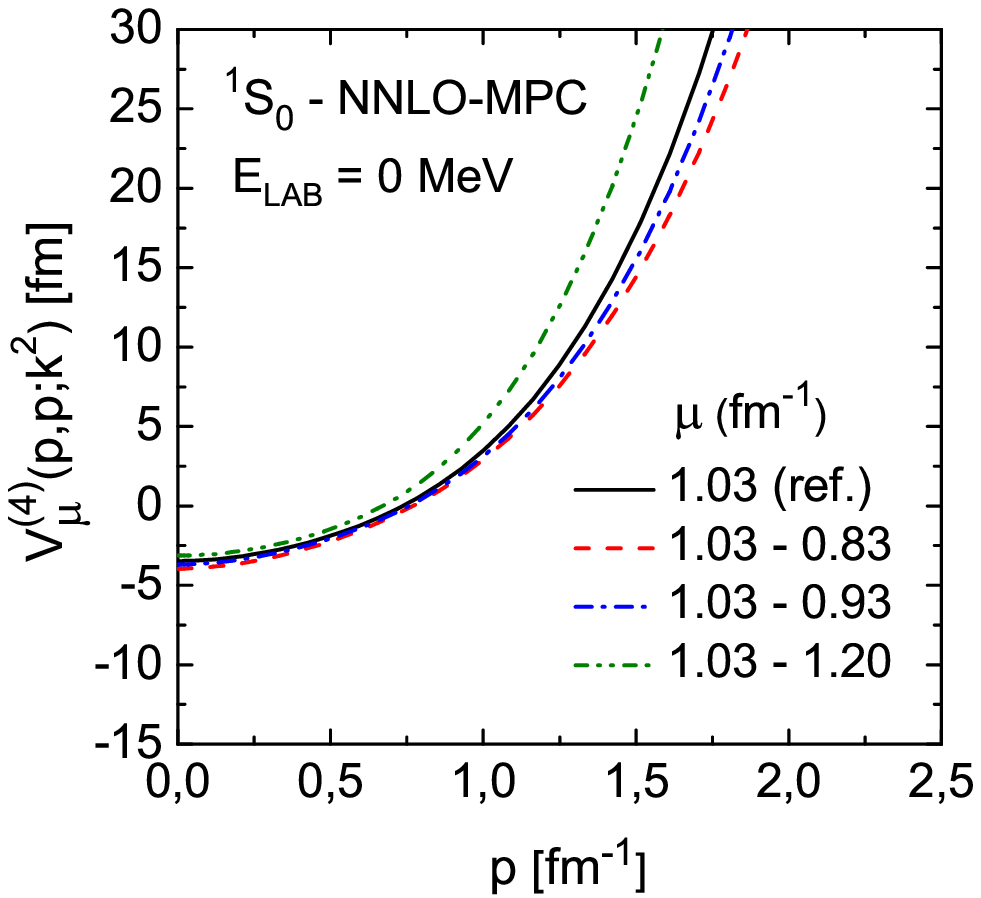}
\includegraphics[scale=0.38]{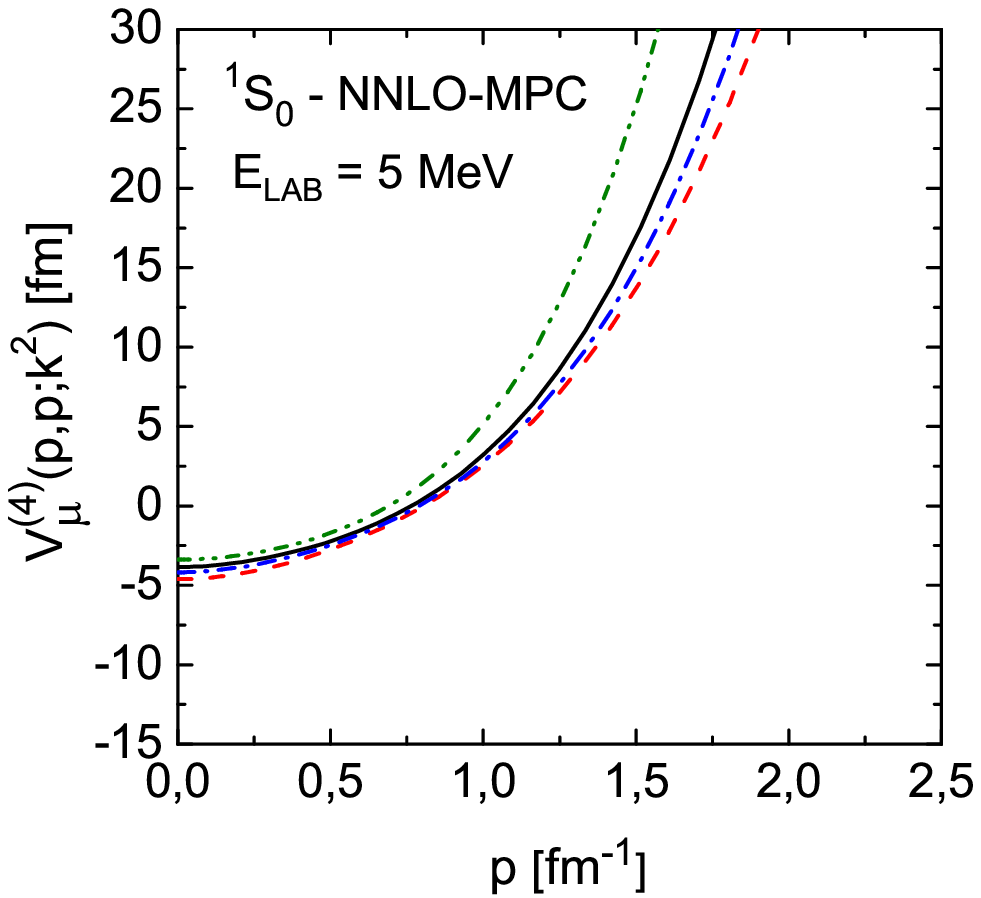}
\includegraphics[scale=0.38]{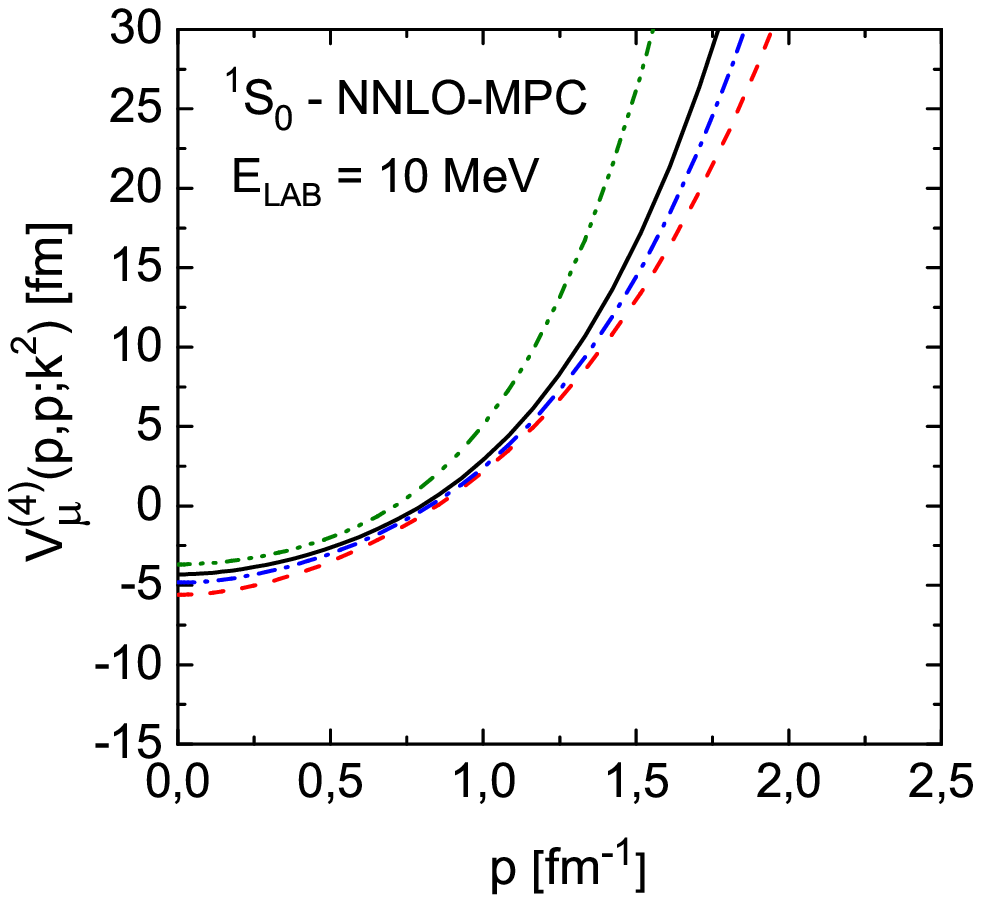}
\includegraphics[scale=0.38]{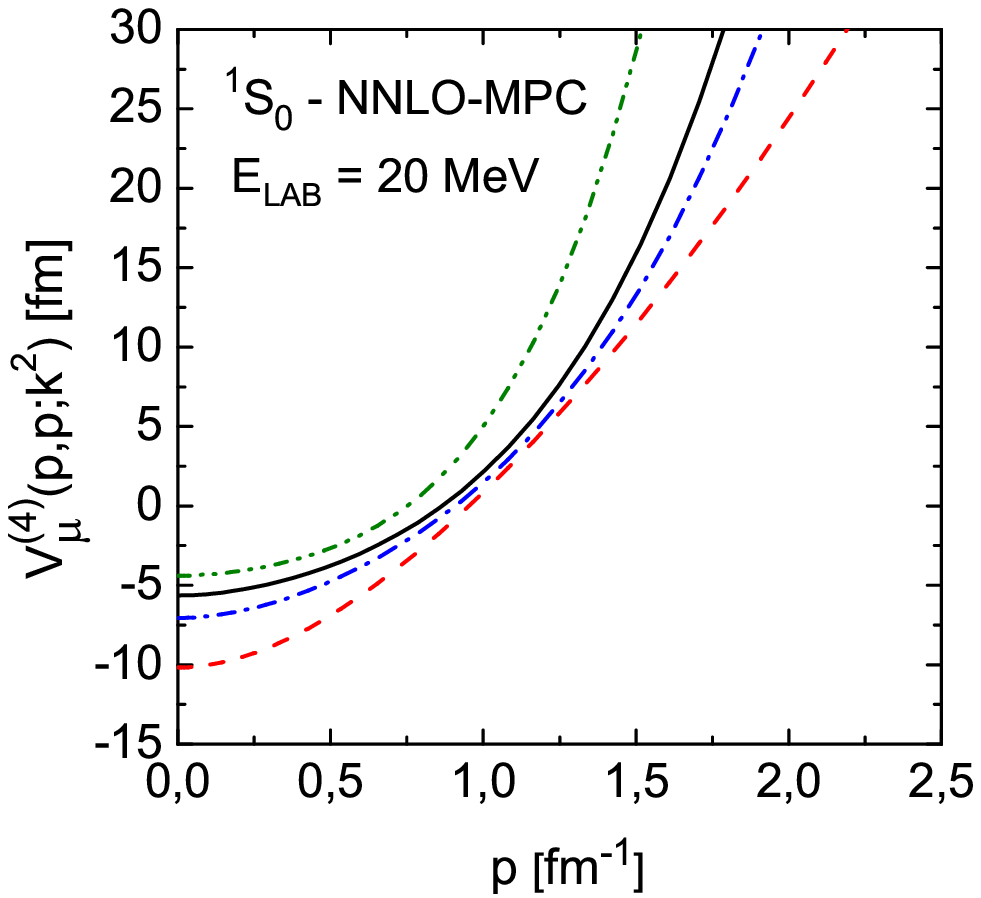}
}
\vspace*{0.2cm}
\centerline{
\includegraphics[scale=0.38]{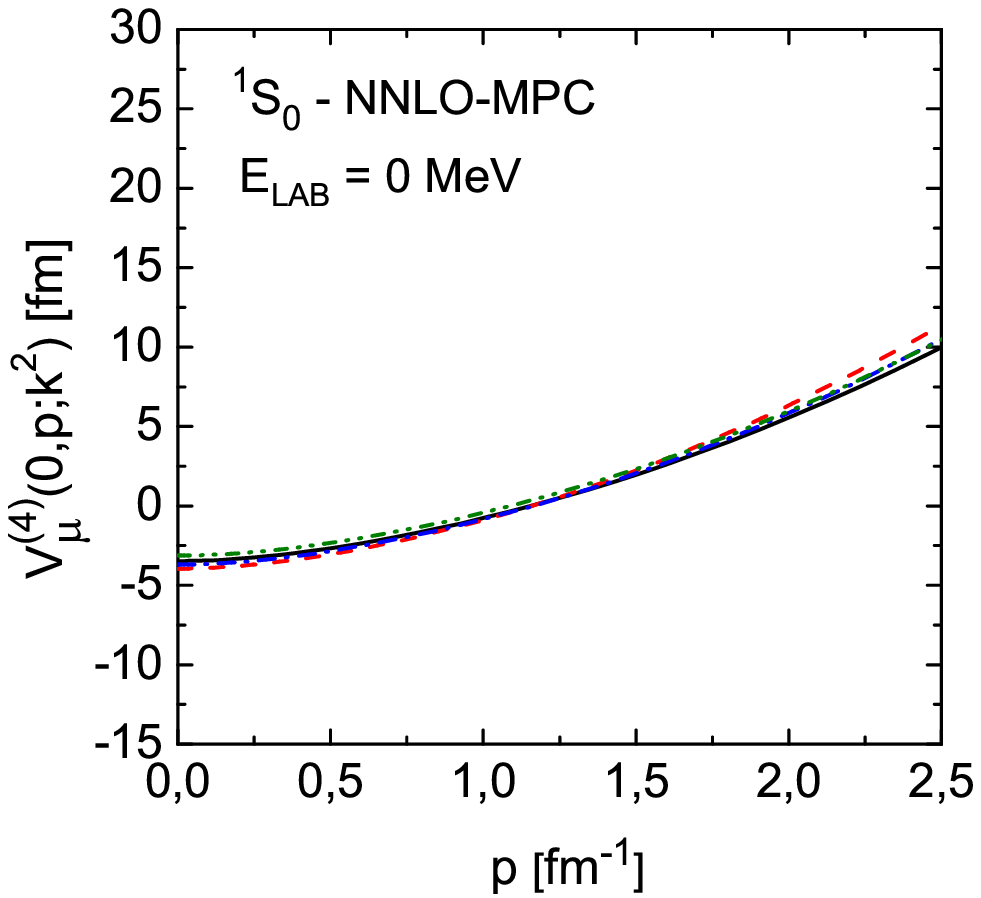}
\includegraphics[scale=0.38]{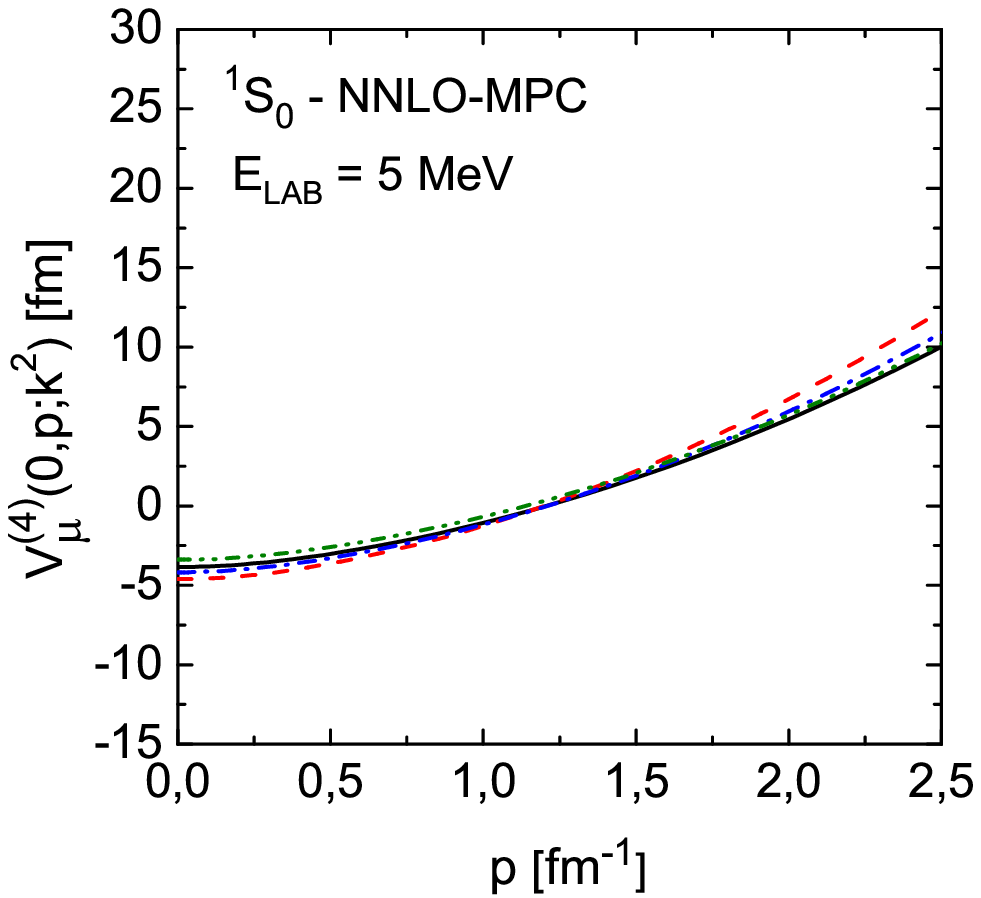}
\includegraphics[scale=0.38]{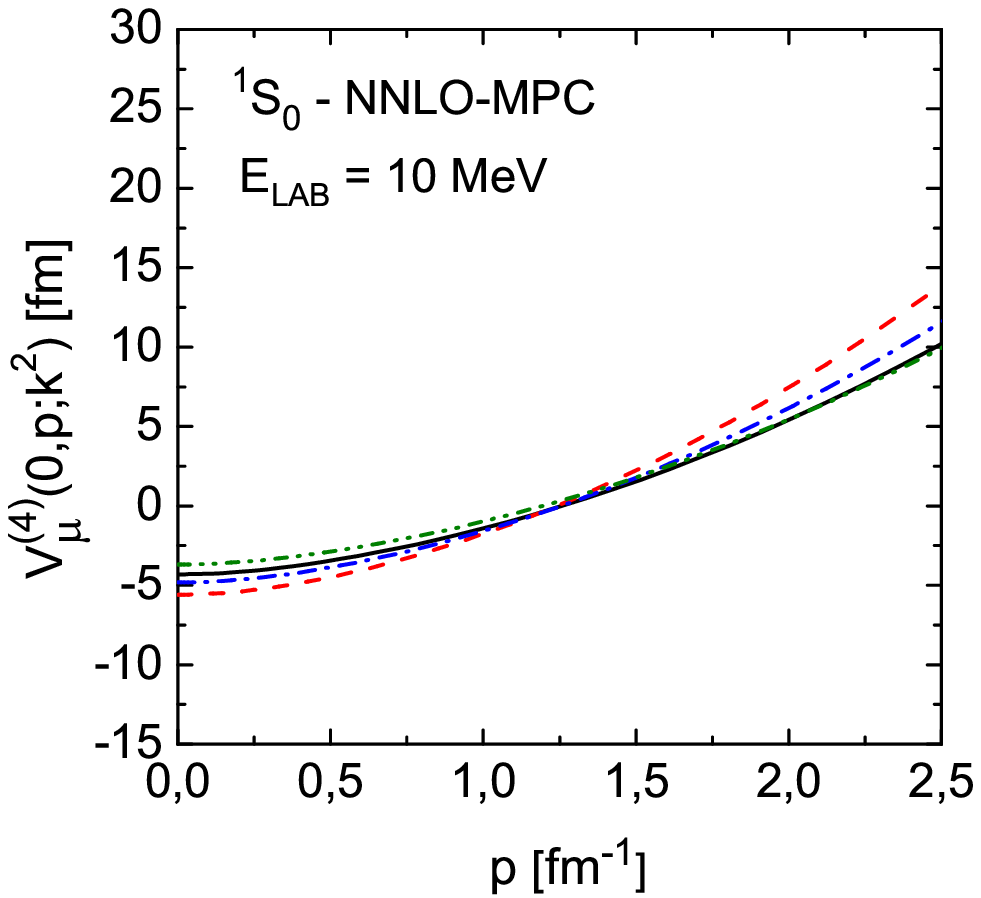}
\includegraphics[scale=0.38]{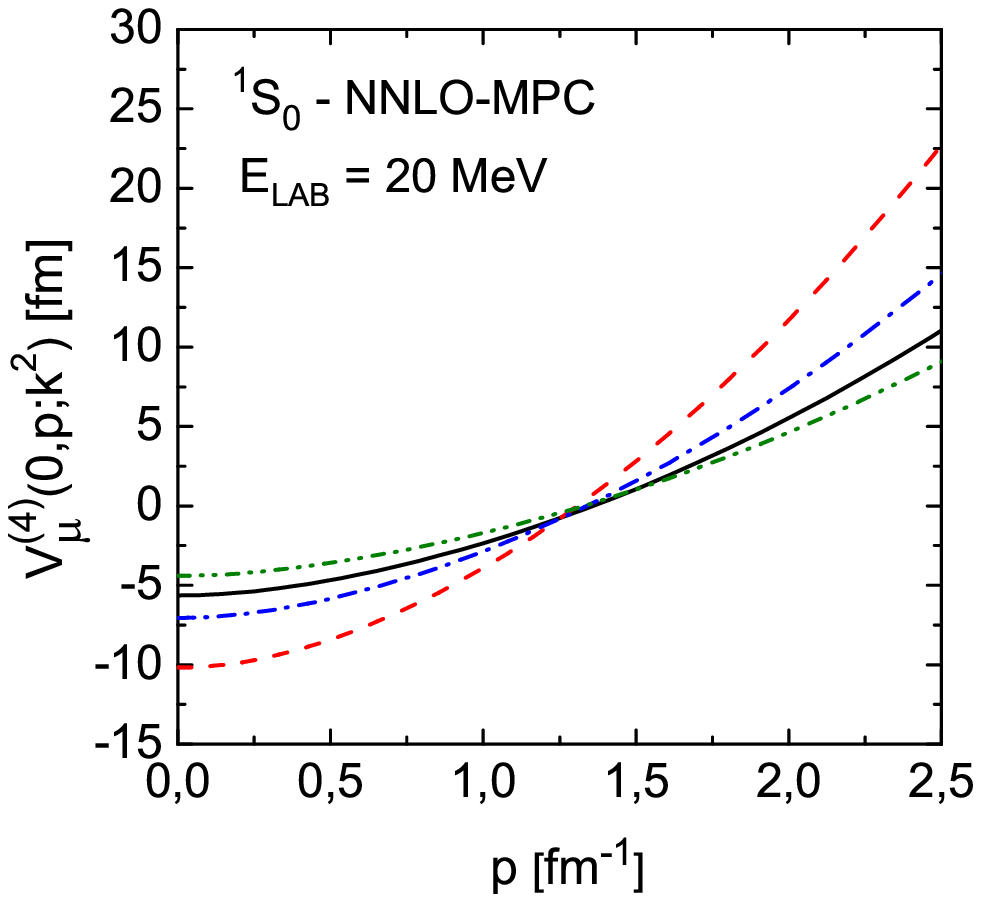}
}
\caption{(Color on-line) Evolution through the NRCS equation of the $^1S_0$ channel driving term $V^{(4)}_{\mu}$ at $NNLO$ in the MPC scheme for several values of $E_{\rm LAB}$. Top panels: diagonal matrix elements; Bottom panels: off-diagonal matrix-elements.}
\label{fig3}
\end{figure*}

In order to explicitly demonstrate the renormalization group invariance in the SKM approach, we consider the evolution through the NRCS flow equation of the $^1S_0$ channel driving term at $NNLO$ in the MPC scheme. In a partial-wave relative momentum space basis, the NRCS flow equation for the matrix-elements of the driving term $V^{(n)}_{\mu}$ in uncoupled channels is given by
\begin{eqnarray}
\frac{\partial V^{(n)}_{\mu}(p,p';k^2)}{\partial\mu^2}&=&\frac 2 \pi \int_0^\infty dq~q^2\left[n\frac{(\mu^2+k^2)^{n-1}}{(\mu^2+q^2)^{n+1}}\right]\nonumber \\&\times&
V^{(n)}_{\mu}(p,q;k^2)~V^{(n)}_{\mu}(q,p';k^2) \; .
\label{CSEnPW}
\end{eqnarray}

We solve Eq. (\ref{CSEnPW}) numerically for $n=4$, obtaining an exact (non-perturbative) solution for the evolved $^1S_0$ channel driving term $V^{(4)}_{\mu}$. The relative momentum space is discretized on a grid of $200$ gaussian integration points, leading to a system of $200 \times 200$ non-linear first-order coupled differential equations which is solved using an adaptative fifth-order Runge-Kutta algorithm. In Fig.~\ref{fig3} we show the evolution of the diagonal matrix-elements (top panels) and the off-diagonal matrix-elements (bottom panels) of the $^1S_0$ channel driving term $V^{(4)}_{\mu}$ from a reference subtraction scale ${\bar \mu}=1.03~\rm{fm}^{-1}$ to $\mu=0.83$, $0.93$ and $1.2~\rm{fm}^{-1}$ for several values of $E_{\rm LAB}$. One should note that the solution of the NRCS flow equation leads to a non-trivial evolution of the driving term $V^{(4)}_{\mu}$ with the sliding subtraction scale $\mu$. The change in the form of the driving term is not only due to the running of the renormalized strengths of the contact interactions $C_i(\mu)$ but also due to the new operators that are generated by the NRCS flow.
\begin{figure}[t]
\centerline{\includegraphics[scale=0.75]{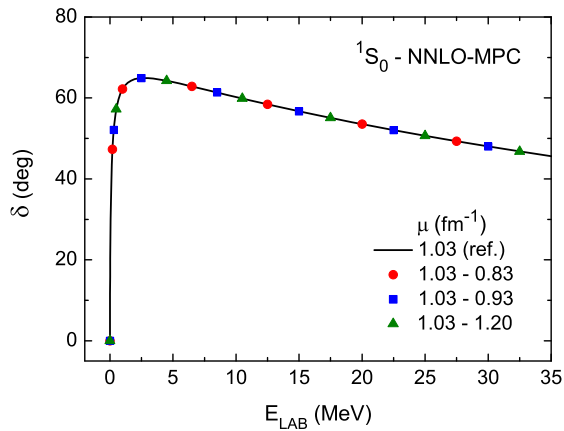}}
\caption{(Color on-line) Phase-shifts in the $^1S_0$ channel obtained with the driving term $V^{(4)}_{\mu}$ at $NNLO$ in the MPC scheme evolved through the NRCS equation. For comparison, we also show the phase-shifts obtained with the driving term $V^{(4)}_{\bar \mu}$ computed at the reference scale ${\bar \mu}$ (solid line).}
\label{fig4}
\end{figure}

As shown in Fig.~\ref{fig4}, the evolution of the $^1S_0$ channel driving term $V^{(4)}_{\mu}$ through the NRCS equation ensures that the phase-shifts calculated from the solution of the LS equation for the $4$-fold subtracted $K$-matrix remain invariant (except for relative differences smaller than $10^{-12}$ due to numerical errors). Remarkably, this result still holds even when the driving term is evolved to subtraction scales $\mu$ which are beyond the limit imposed by the Wigner causality bound (such as $\mu=1.2~\rm{fm}^{-1}$). This can be understood in the following way. As pointed before, the input physical information (e.g. the experimental values of the ERE parameters $a$, $r_e$, $v_2$, etc) is encoded in the initial driving term $V^{(n)}_{\bar{\mu}}$, which is determined by fixing the renormalized strengths $C_i({\bar \mu})$ of the contact interactions at the reference subtraction scale $\bar{\mu}$ to fit data and sets the boundary condition for the NRCS evolution. Thus, once the constraint imposed by Wigner's bound is fulfilled at the boundary condition, provided one chooses a reference scale $\bar{\mu} < \mu_{max}$ (which in this case is $\sim 1.1~\rm{fm}^{-1}$), the NRCS flow equation, Eq. (\ref{CSEnPW}), can be solved to evolve the driving term to any subtraction scale $\mu$ while keeping it hermitian. These results clearly show that the sliding subtraction scale $\mu$ vanishes as a physical parameter, the only relevant scale being the reference scale $\bar{\mu}$ where the boundary condition for the NRCS evolution is determined.

\section{Summary and conclusions.}

The subtracted kernel method (SKM) approach presented in this work provides a powerful renormalization group invariant method to renormalize singular two-body interactions. The iterative procedure involving recursive multiple subtractions performed in the kernel of the scattering equation allows for a systematic treatment of chiral effective field theory (ChEFT) nucleon-nucleon ($NN$) potentials up to higher-orders in the chiral expansion, which include pion-exchange and contact interactions that present increasingly strong singularities at short-distances. This also makes the SKM approach a convenient tool to implement power-counting schemes in which higher-order contact interactions are promoted to lower-order, since ultraviolet (UV) power divergences of any order can be properly handled by performing a number of subtractions enough to render a finite amplitude. A disadvantage of the SKM formalism is that the computational load increases as more subtractions are performed: in order to compute the $n$-fold subtracted amplitude, the iterative procedure requires $n$ matrix inversions.

We have shown that a modified power counting scheme (MPC) based on the promotion of contact interactions, such that at each order in the chiral expansion one new contact interaction is included in the effective $NN$ potential, yields a systematic order-by-order power-law improvement in the $^1S_0$ channel phase-shifts calculated up to next-to-next-to-leading-order ($NNLO$), that is lacking in Weinberg's power counting scheme (WPC). Our results show that the scaling of the relative errors at low-energies is nearly dominated by the tuning of the renormalized strengths of the contact interactions, and so suggest that the short-range contact interactions are relatively more important than the long-range pion-exchange interactions. An essential ingredient in our calculations is the procedure employed to fix the renormalized strengths of the contact interactions, based on the fitting of ``data" generated from the Nijmegen II potential for a spread of very low on-shell momenta, which allows for a clear analysis of the error scaling of predicted phase-shifts, and hence of the power counting, using ``Lepage plots". Such a procedure is equivalent to a fitting of the experimental values of the effective range expansion (ERE) parameters ($a$, $r_e$, $v_2$, etc) to a given order in $k^2$, where $k$ is the on-shell momentum in the center-of-mass frame.

One should note that we are not analyzing the power counting at the level of the chiral expansion for the $NN$ potential, but ``a posteriori" at the level of the observables, i.e. after iterating the potential in the subtracted Lippmann-Schwinger (LS) equation. The lack of power-law improvement in the $^1S_0$ channel phase-shifts we have observed at $NNLO$ in the WPC scheme shows that the corresponding subtracted amplitude do not follow the power counting of the chiral $NN$ potential, which is the motivation for our MPC scheme. As pointed out by several works \cite{lepage,gegelia2,valderrama6,valderrama7,valderrama8,yang2,yang3}, a consequence of the singular nature of the chiral $NN$ interactions at short-distances is that the full iteration of the potential in the scattering equation can change the scaling behavior of each of the interactions included, and hence their relative importance to the amplitude, thus leading to modifications of the power counting at the level of the observables.

We have also demonstrated by explicit numerical calculations for the scattering of two-nucleons in the $^1S_0$ channel that the SKM procedure applied to the chiral $NN$ interactions up to $NNLO$ is renormalization group invariant under the change of the subtraction scale. Once the renormalized strengths of the contact interactions are fixed at a reference scale to fit low-energy scattering observables, the subtraction scale can be changed by evolving the driving term of the subtracted LS equation through a non-relativistic Callan-Symanzik (NRCS) flow equation, such that the results for the calculated phase-shifts remain invariant. Moreover, the sliding subtraction scale to which the driving term can be evolved is not constrained by the Wigner causality bound. In this way, the sliding subtraction scale vanishes as a physical parameter. The relevant scale parameter left in the theory is the reference scale where the boundary condition of the NRCS flow equation is determined through the input of physical information, which can then be regarded as a renormalization scale.

It is important to emphasize that in order to properly assess the effectiveness of the MPC scheme we have described in this work, as well as the renormalization group invariance in the SKM approach, it is necessary to carry out the calculations up to next-to-next-to-next-to-leading-order ($N^3LO$) and perform a comprehensive analysis for all partial-wave channels relevant at low-energies (e.g. for total angular momentum $j \leq 5$). In particular, the calculations for the lower partial-waves would be essential to address the consequences of our results for applications of ChEFT to few- and many-nucleon systems (e.g. light-nuclei, and nuclear and neutron matter). If we do find, for all relevant partial-wave channels and up to $N^3LO$, that a systematic order-by-order power-law improvement can be achieved through the MPC scheme and the $NN$ phase-shifts remain invariant under the change of the subtraction scale through the NRCS flow equation, then it would be reasonable to expect that similar results may also be obtained when applying the SKM approach to few- and many-nucleon systems. Of course this must be verified through explicit calculations, which would require the extension of the SKM approach to solve dynamical equations pertinent to nuclear few- and many-body problems (e.g. the Faddeev-Yakubovsky equations \cite{nogga2} and the Bethe-Brueckner-Goldstone equations \cite{machleidt4}). Nevertheless, due to its sensitivity to the short-range contributions to the chiral $NN$ potential \cite{deltaiso5,valderrama2}, the $^1S_0$ channel provides a good starting point to investigate the interplay between the pion-exchange and the contact interactions in the non-perturbative renormalization of $NN$ interactions in ChEFT.

In forthcoming works, we will extend the calculations presented here to other partial-wave channels and up to $N^3LO$, aiming to perform a systematic analysis of the power counting and the renormalization group invariance in the SKM approach for chiral $NN$ interactions. In particular, we will investigate in detail the impact of the long-range pion-exchange interactions and their interplay with the short-range contact interactions. We also intend to compare the evolution of chiral $NN$ interactions through the NRCS equation in the SKM approach with the evolution through Birse's renormalization group (RG) equation in cutoff EFT, both in a perturbative and in a non-perturbative renormalization framework.

Furthermore, we want to investigate the possibility of generalizing the SKM approach to renormalize ChEFT potentials including three- and four-nucleon interactions. We believe that this generalization can be worked out by using techniques similar to those developed in the context of EFT for three-body systems with contact interactions \cite{subren1,subren2,subren3,subren4}, which allow to obtain renormalized amplitudes by performing subtractions in the three-body scattering equations. By accomplishing such a generalization, we expect to be able to apply the SKM approach in sensible calculations of few- and many-nucleon systems.

\ack{S.S. was supported by Instituto Presbiteriano Mackenzie through Fundo Mackenzie de Pesquisa and FAPESP and V.S.T. by FAEPEX/PRP/UNICAMP and FAPESP. Computational power provided by FAPESP grants 2011/18211-2 and 2010/50646-6.}

\end{document}